\documentclass[aps,prl,reprint,groupedaddress]{revtex4-1}

\usepackage{amsmath}
\usepackage{graphicx}
\usepackage{upgreek}
\usepackage{graphicx}
\usepackage{subfig}
\usepackage{caption}
\begin{document}

\title{Downramp-assisted underdense photocathode electron bunch generation\\ in plasma wakefield accelerators}

\author{A. Knetsch$^1$, O. Karger$^1$, G. Wittig$^1$, H. Groth$^1$, Y. Xi$^2$, A. Deng$^2$, J.B. Rosenzweig$^2$, D.L. Bruhwiler$^{3,4}$, J. Smith$^5$, D.A. Jaroszynski$^6$, Z.-M. Sheng$^6$, G.G. Manahan$^6$, G. Xia$^7$, S.P. Jamison$^8$, B. Hidding$^{1,2,6}$}

\affiliation{$^1$Institute for Experimental Physics \& CFEL, University of Hamburg, Germany, \\$^2$Particle Beam Physics Laboratory, UCLA, USA, $^3$RadiaSoft LLC, USA, $^4$RadiaBeam Technologies LLC, USA, \\$^5$Tech-X UK Ltd, Warrington, UK \\$^6$SUPA, Department of Physics, University of Strathclyde, UK, $^7$University of Manchester, Manchester, UK, $^8$STFC/DL/ASTeC/Cockcroft Institute, Daresbury, UK\\
\\
\textnormal{submitted}}

\begin{abstract}
It is shown that the requirements for high quality electron bunch generation and trapping from an underdense photocathode in plasma wakefield accelerators can be substantially relaxed through localizing it on a plasma  density downramp.  This depresses the phase velocity of the accelerating electric field until the generated electrons are in phase, allowing for trapping in shallow trapping potentials. As a consequence the underdense photocathode technique is applicable by a much larger number of accelerator facilities. Furthermore, dark current generation is effectively suppressed.
\end{abstract}


\maketitle
Plasma wakefield acceleration utilizes the collective electric field arising from transient separation of electrons and ions in plasma \citep{AkhiezerFainberg1949,Fainberg1956,Veksler1956,Budker1956}, which using the classical wave-breaking limit \cite{Dawson1959PhysRev.113.383,AkhiezerPolovinJETP1956} can be approximated as $E_\mathrm{wb} \mathrm{[V/m]} \simeq 96\,\sqrt{{n_\mathrm{e}}  \mathrm{[cm^{-3}]}}$ \cite{Esarey1996}. Thus electron densities of $n_\mathrm{e} \approx 10^{17}$ $\mathrm{cm^{-3}}$ can yield electric fields of $E\approx 30$ GV/m. The highest electron energies produced in linear accelerators have been achieved using electron beam driven plasma wakefield acceleration (PWFA) at SLAC \cite{Blumenfeld2007,Litos2014Natureshort}. In addition to the orders of magnitude higher energy gains in plasma accelerators, compared with classical radio frequency driven metallic cavities, the plasma blowout cavities are also an order of magnitude smaller, with longitudinal sizes being approximately equal to the plasma wavelength $\lambda_\mathrm{p} =2 \pi c/\omega_\mathrm{p} \approx 100\ \mu\mathrm{m}$, which depends on the plasma frequency  $\omega_\mathrm{p}=\sqrt{n_\mathrm{e} e^2 /(m_0 \epsilon_0)}$, where $e$ and $m_0$ are the electron charge and rest mass, respectively, and $\epsilon_0$ is the vacuum permeability. Electron bunches generated and/or accelerated within such plasma blowouts are also intrinsically ultrashort, which is another advantageous feature of plasma accelerators. However, the obtainable electron beam quality in terms of emittance, brightness, energy spread, and the controllability and tunability of PWFA systems have until recently been considered to be somewhat limited. 

A novel approach known as  Trojan Horse underdense photocathode PWFA, where an electron beam drives
the plasma blowout in a low-ionization threshold (LIT) medium plasma, and independently electrons are released inside this plasma blowout by local ionization of an underdense high ionization threshold (HIT) medium with a strongly focused and synchronized laser pulse \cite{hiddingpatent2011short,Hidding2012,hidding570beyondshort,YunfengPhysRevSTABshort,LiPRL2013short,PhysRevLett.112.035003Xu2014}, promises to exceed in many ways the electron bunch quality and tunability of even the best traditional photocathode-based accelerator systems. 
PWFA using electron bunches as plasma wave drivers \cite{Chen1985PhysRevLettshort,Rosenzweig1988PRLshort}  is attractive because of the much longer possible acceleration lengths when compared with laser-plasma wakefield acceleration (LWFA) \cite{Tajima1979,Pukhov2002}: the effects of dephasing and transverse drive beam expansion are much smaller because of the high Lorentz factor $\gamma_\mathrm{d} = 1 + W/m_0 c^2$ associated with the electron driver beam of energy $W$ in PWFA. In contrast, in LWFA the plasma wave phase velocity $v_\mathrm{ph}$, which is equal to the laser pulse group velocity $v_\mathrm{g,l}$, depends on the plasma frequency $\omega_\mathrm{p}$ and the laser frequency $\omega_0$ because $v_\mathrm{ph} \approx v_\mathrm{g,l} = c(1-\omega_\mathrm{p}^2/\omega_0^2)^{1/2}$. Thus for typical plasma densities $\gamma_\mathrm{ph}=[1-(v_\mathrm{ph}/c)^2]^{-1/2} \approx 130$ is much lower than the usual drive bunch $\gamma_\mathrm{d}$ used in PWFA.  
While the higher $\gamma$ and the rather elliptic blowout shapes in PWFA result in reduction of unwanted self injection and dark current compared with LWFA and its more spherical bubble shapes \cite{KostyukovTrappingPRL2009short}, on the other hand the desired high-quality witness bunch generation and trapping is more difficult in the first place. One has to produce multi-kA drive beam currents required to excite an electrostatic wake potential  $\Psi$ that exceeds the trapping threshold\cite{Kirby2009PRSTshort} -- which is challenging even if electrons can be released in the center of the plasma wave at the minimum trapping potential via the underdense photocathode mechanism.  Therefore, to date, only one facility has the prerequisites to experimentally demonstrate underdense photocathodes -- FACET at SLAC with its driver beam current $I_\mathrm{d} > 10\ \mathrm{kA}$, approaching the Alfv\'{e}n current $I_\mathrm{A} \approx 17\ \mathrm{kA}$. 
To make the scheme widely accessible, it is either necessary to increase the peak currents of other state-of-the-art electron accelerators, e.g. by advanced bunch generation and compression schemes, or by searching for alternative techniques to facilitate trapping.  

In this Letter, it is shown that the requirements of the driving electron beam are substantially reduced by using an underdense photocathode situated on a plasma density downramp (Downramp Trojan Horse, DTH), which makes the generation of tunable electron witness bunches with ultrahigh brightness $B \approx 10^{19} \mathrm{A m^{-2} rad^{-2}}$ accessible to a wider range of accelerators, including beams from LWFAs. Density downramps are well-known  in LWFA 
\cite{BulanovDensityTransition1998PhysRevEshort,GeddesDensityDownrampPRL2008short,SchmidDensityTransPhysRevSTABshort,GonsalvesNatPhys2011short}, and have early been proposed for PWFA \cite{SukRosenzweigDensityTransition2001PhysRevLettshort}. 
Here, we explore a transitional regime, where in the first plasma wave bucket neither density downramp self-injection occurs (which would produce unwanted dark current) nor photocathode-released electrons would be trapped in absence of the downramp -- but laser-released electrons are trapped in presence of a downramp.

As a concrete example, we consider a $Q = 150$ pC, $W= 250\ \mathrm{MeV}$ (i.e. $\gamma_\mathrm{d} \approx 500$) electron driver bunch with a peak current of $ I_\mathrm{d} \approx 3\ \mathrm{kA}$, Gaussian shape with length of $\sigma_z = 6\ \mu \mathrm{m}$ rms and transverse size of $\sigma_r = 7\ \mu\mathrm{m}$ rms, and an energy spread of \mbox{$\Delta W/W = 0.02$}. Electron bunches with these characteristics are available from a range of accelerators, including laser-plasma-accelerators \cite{Wiggins1percent2010PPCFshort,LundhKiloampsNatPhys2011short}. 
3D particle-in-cell (PIC) simulations using VSim/VORPAL \cite{Nieter2004448} show that such a bunch effectively drives a blowout cavity in preionized hydrogen (the LIT component) of electron density \mbox{$n_\mathrm{H} = 1.5 \times 10^{17}\ \mathrm{cm^{-3}}$}, corresponding to a plasma wavelength of $\lambda_\mathrm{p} \approx 86\ \mu\mathrm{m}$. 
The HIT component is helium, which has the same density. In contrast to related laser-driven schemes \cite{Umstadterpatent1995,Umstadter1996CollidingPulsesPhysRevLett.76.2073,ChenJAP200610short}, when electron beams are used as drivers, they offer the significant advantage that the plasma blowout cavity  is produced by the repulsive force of the unipolar electric field of the electron beam instead of the gradient of the laser intensity $I\propto E^2$. Although only producing a peak electric self field of $E_{\mathrm{max}}(r) \approx Q\{1-\exp[-r^2/(2\sigma_r^2)]\}/(2\pi^{3/2}\epsilon_0\sigma_z r) \approx 16.5\ \mathrm{GV/m}$ (confirmed by the PIC simulations), the drive beam does excite the H-plasma blowout, but He remains neutral because the  fields are well below the tunnel ionization threshold. The synchronized underdense photocathode laser pulse is set to a normalized amplitude of the laser vector potential $a_0 = 0.015 \ll 1$, corresponding to a peak electric field of $E_0 \approx 60.2\ \mathrm{GV/m}$, which is high enough to ionize He at the focus, but without imparting significant residual transverse momentum to the He electrons which would alter their trajectories and spoil their emittance. The laser pulse of rms duration $\tau = 40\ \mathrm{fs}$ strongly focuses to a beam waist of $w_0=4\, \mu\mathrm{m}$ as it co-propagates with the electron beam driver, which reduces the phase space volume of the generated bunch. 

The laser pulse is implemented as a TEM$_{00}$ Gaussian beam and defined in envelope approximation, i.e.
$E_{x}=E_0[w_0/w(z)]\exp\{-[r/w(z)]^2\}\exp\{-[t^2/(2\tau^2)]\}\mathrm{,}$ where $w(z)=w_0[1+(z/z_0)^2]^{1/2}$ is the beam waist at longitudinal position $z$ and $z_0=\pi w_0^2/\lambda$ is the Rayleigh length. The laser pulse releases electrons at $\xi=z-ct = -38\ \mu\mathrm{m} \approx \lambda_\mathrm{p}/2$ behind the center of the electron beam, which is around the minimum of the wake potential, i.e. at the zero-crossing where the longitudinal plasma wakefields are minimal.  
Such a laser pulse electric field superimposed on the blowout electric plasma field (taken from the simulation) is estimated to yield a total helium electron charge of $Q \approx 25.3\ \mathrm{pC}$, by numerically integrating over the ADK tunnelling ionization rates as described in \cite{Bruhwiler2003short}.
In this $k_\mathrm{p} \sigma_z<\sqrt{2}$ scenario, the dimensionless beam charge $\tilde{Q} = N_\mathrm{b} k_\mathrm{p}^3/n_\mathrm{e} \approx 1.84$, relating the total number of bunch electrons $N_b$ to the plasma electron number inside a spherical volume with radius of the skin depth $k_\mathrm{p}^{-1}$, indicates that the blowout is not strongly nonlinear \cite{BarovEnergyLossQtildePRST2004short}, which corresponds to a moderately intense electron bunch driver.    
Trapping of laser-released electrons in the blowout region is not possible in this case, which is confirmed by PIC-simulations that are discussed later. 

A density downramp in the plasma density $n_\mathrm{e}$ increases the plasma wavelength $\lambda_\mathrm{p}\propto 1/\sqrt{n_\mathrm{e}}$ and therefore the blowout length accordingly. The density transition has to be chosen sufficiently sharp  to enable  trapping of electrons released in the blowout center by the underdense photocathode laser, while avoiding  traditional downramp injection at the end of the blowout cavity. According to \cite{SukRosenzweigDensityTransition2001PhysRevLettshort}, the latter type of (here unwanted) trapping is suppressed if the downramp length $L$ is longer than the plasma skin depth $k_\mathrm{p}^{-1}=c/\omega_\mathrm{p}$. 
 As the blowout region expands longitudinally, its phase velocity retards to $v_\mathrm{ph}=c\Big(1+1/2n_\mathrm{e}(z)\frac{\partial n_\mathrm{e}(z)}{\partial z}\xi \Big)^{-1}$ \cite{PhysRevEFubianiDensityGradientsshort} at position $\xi=z-ct$ relative to the center of the  driving electron beam in the co-moving frame, where $n_\mathrm{e}(z)$ is the electron density at the longitudinal position $z$ in the laboratory frame. 
Because $v_\mathrm{ph}$ is decreased on the downramp, electrons released inside the blowout can catch up with the driver beam's velocity more easily, which facilitates trapping.

\begin{figure}[htbp]
\includegraphics[width=0.45\textwidth]{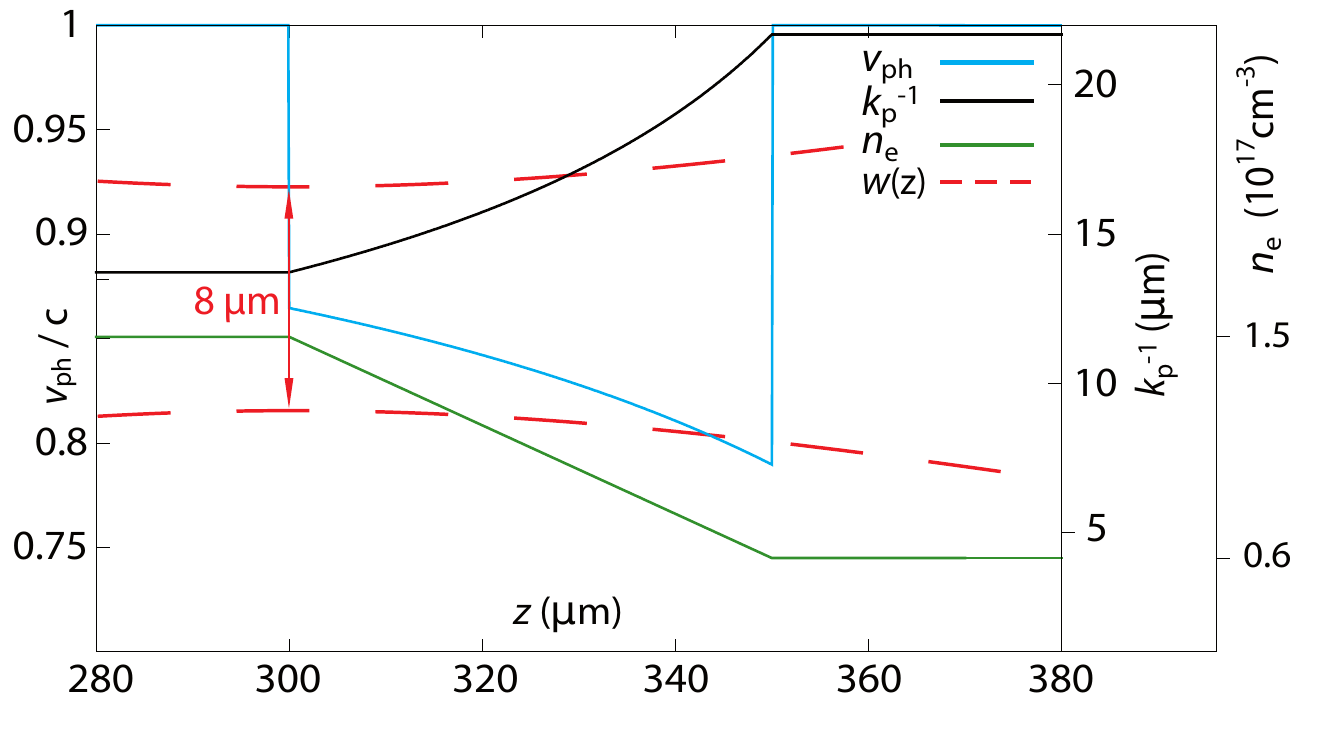}
\caption{Electron density profile $n_\mathrm{e}$ of the downramp (green), plasma skin depth $k_\mathrm{p}^{-1}$ (black) and phase velocity $v_\mathrm{ph}$ (blue) at $\xi=-38\ \mu\mathrm{m}$ behind the driver.
The ionizing photocathode laser pulse transverse size $w(z)$ with its focus at $z=300\ \mu\mathrm{m}$ (dashed line) is also shown, using the same $y$-axis as with $k_\mathrm{p}^{-1}$.}
\label{fig:phase_velocity}
\end{figure}

Figure \ref{fig:phase_velocity} illustrates the scenario by plotting the longitudinal profile of the relevant parameters: electron density $n_\mathrm{e}$ (green line), the corresponding $k_\mathrm{p}^{-1}$ (black),  and $v_\mathrm{ph}$ (blue) at $\xi=-38\ \mu$m behind the bunch driver center. The laser transverse width $w(z)$ is also shown (using the $k_\mathrm{p}^{-1}$ axis scale, dashed line), with its focal plane located at $z=300\ \mu\mathrm{m}$, where also the downramp starts. This way the trapping facilitation effect is maximized for the majority of the released electrons. The plasma density ramp length $L = 50\ \mu\mathrm{m}$  fulfills the condition $L> k_\mathrm{p}^{-1}$ everywhere, to avoid density transition injection \cite{SukRosenzweigDensityTransition2001PhysRevLettshort}, while the initial density $n_\mathrm{i}=1.5\times 10^{17}$ $\mathrm{cm}^{-3}$ decreases linearly to the final density $n_\mathrm{f}=0.4\ n_\mathrm{i}\approx 0.6 \times 10^{17}$ $\mathrm{cm}^{-3}$. 
Experimental realization of such a downramp is possible taking into account that similarly sharp downramps have been experimentally demonstrated \cite{SchmidDensityTransPhysRevSTABshort,BuckShockFrontInjector2013short}.
At the co-moving laser pulse position $\xi=-38$ $\mu\mathrm{m}$ behind the driver electrons, i.e. the birth place of He electrons, as  $k_\mathrm{p}^{-1}$ increases from $k_\mathrm{p}^{-1}\approx 14\ \mu\mathrm{m}$ to $k_\mathrm{p}^{-1}\approx 22\ \mu\mathrm{m}$, the phase velocity $v_\mathrm{ph}$ decreases to a minimum phase velocity of $0.79\,c$ at the end of the ramp, then bounces back to $\beta=1$ when the flat top $n_\mathrm{f}$ is reached.
 \begin{figure}[htbp]
 \hspace*{-0.03\textwidth}
 \includegraphics[width=0.51\textwidth]{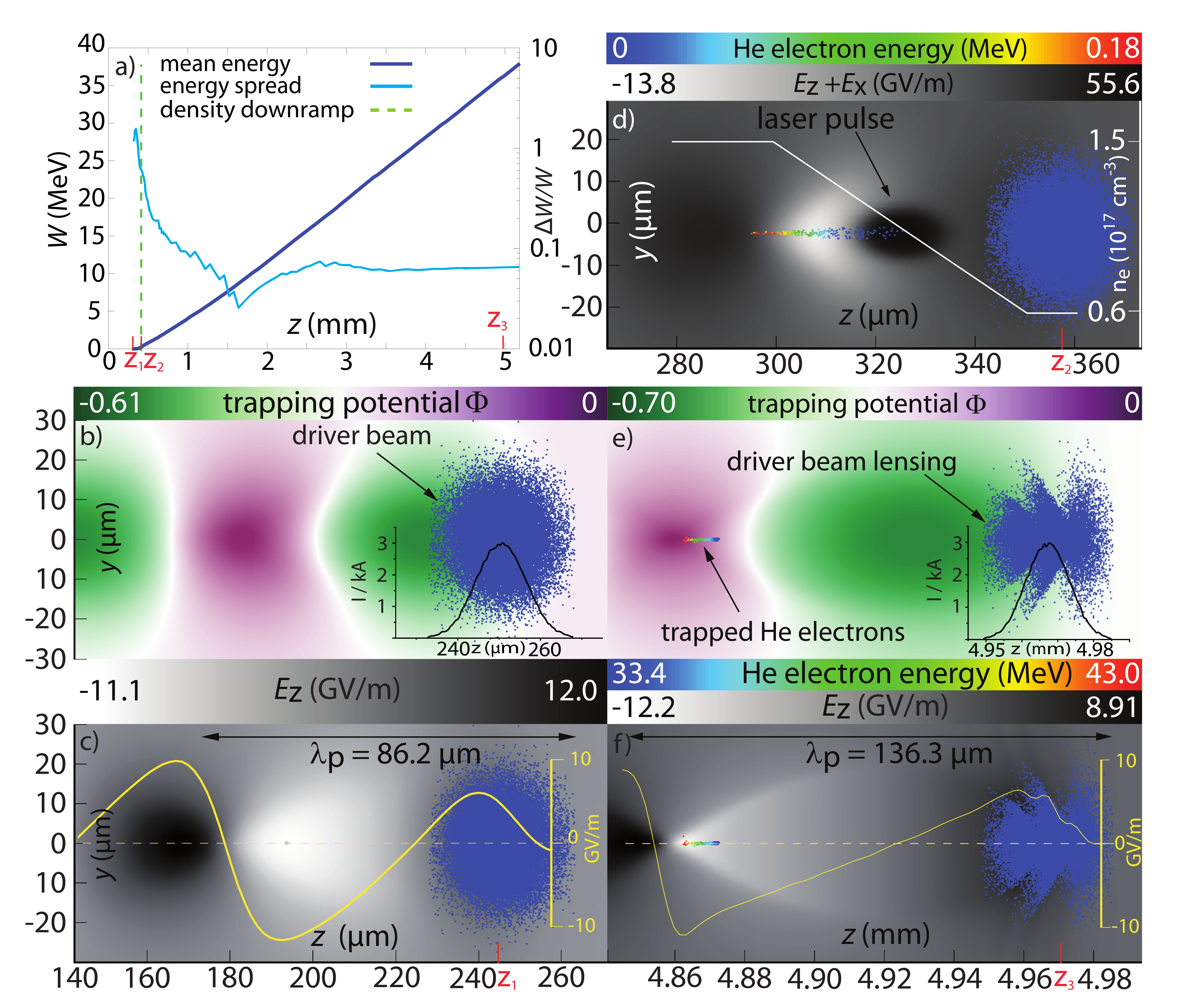}
\caption{PIC simulation results, showing energy gain and energy spread over $z \approx 5$ mm of acceleration (a), trapping potential $\Phi$ (b) and electric field $E_z$ (c) before downramp, photocathode He electron release on downramp (d), and $\Phi$ (e) and  $E_z$ (f) after the downramp, showing the trapped He electron bunch.}
 \label{fig:VISIT}
 \end{figure}
The addition of the described downramp enables trapping despite of the moderate drive beam current. PIC simulations show in figure \ref{fig:VISIT} a), how the released He electrons are trapped and gain energy linearly in the co-propagating blowout, reaching $W \approx 35\ \mathrm{MeV}$ after $z \approx 5\ \mathrm{mm} $ (left $y$-axis) of acceleration, corresponding to a mean gradient of $E \approx 7.8\ \mathrm{GV/m}$. The green dashed vertical line indicates the position of the downramp, and the positions $z_1$ to $z_3$ mark the situation before the downramp (figure \ref{fig:VISIT} b) and c), on the downramp (figure \ref{fig:VISIT} d) and after $z \approx 5\ \mathrm{mm} $ (figure \ref{fig:VISIT} e) and f). The energy spread  saturates at $\Delta W/W \approx 0.07$ after $z \approx 3\ \mathrm{mm}$ of acceleration (right $y$-axis), which is due to partial He electron loss as discussed below. 
Figure \ref{fig:VISIT} b) shows the trapping potential $\Phi=[\Psi_\mathrm{max}-\Psi]/[(m_\mathrm{0} c^2 /e)(1- \gamma_\mathrm{ph}^{-1})]$ around position $z_1$ just before the downramp, calculated by numerically integrating  the longitudinal electrostatic wake potential $\Psi$ from the electric field derived from the PIC simulations.  The electron bunch drive beam is shown in a 2D projection of the drive beam macro-particles with a longitudinal line-out of the drive beam current, peaking at $I_\mathrm{d}\approx 3\ \mathrm{kA} $. A trapping potential of $\Phi<-1$ indicates that an electron released at rest at this position can be trapped in the blowout region \cite{PakPRL2010PhysRevLettshort}.
Here, the minimum trapping potential is only $\Phi\approx -0.61$ for $n_\mathrm{e}=n_\mathrm{i}$ and $\Phi\approx -0.70$  for $n_\mathrm{e}=n_\mathrm{f}$. Therefore, no trapping is expected without a downramp, which is confirmed by separate simulations with flat plasma density profiles (see figure 3 a). 
In figure \ref{fig:VISIT} c), the longitudinal electric field corresponding to the potential $\Phi$ is represented by a color plot and an on-axis line-out. The field maximum is $E_z \approx 12\ \mathrm{GV/m}$. 
Figure \ref{fig:VISIT} d) shows the underdense photocathode in action: As the laser pulse reaches its focus at $z = 300\ \mu\mathrm{m}$ it releases He electrons, which begin to gain energy. To visualize the laser pulse, which is polarized in the figure plane, the sum of the longitudinal $E_z$ and transverse $E_x$ fields is shown.  

 Experimentally, positioning the laser pulse focus at the start of the plasma density downramp is straightforward, as $\mu\mathrm{m}$-level spatial adjustment is routinely achievable in state-of-the-art experiments. 
After the downramp and the reduction of H-plasma density, the plasma wavelength is $\lambda_\mathrm{p} \approx 136 \ \mu\mathrm{m}$, and figure \ref{fig:VISIT} e) and f) show the trapping potential  $\Phi$ and the longitudinal $E$-field, respectively, after $z \approx 5\ \mathrm{mm}$ of acceleration at this density level. The drive beam shows an increase in the normalized emittance from $\epsilon_{\mathrm{n,rms}} \approx 2.3\times 10^{-6}\ \mathrm{m\ rad}$ to $ \epsilon_{\mathrm{n,rms}} \approx 9.1\times 10^{-6}\ \mathrm{m\ rad}$ and a strong signature of underdense plasma lensing and betatron oscillations  with the betatron function defined as $\beta = \gamma \sigma_r^2 /\epsilon_\mathrm{n}$.
While the longitudinal drive bunch current is unaffected (compare lineouts in \ref{fig:VISIT} b) and e), these effects have a significant impact on the local drive bunch density $n_b$ and consequently also on $\tilde{Q}$. The increase of $\tilde{Q}$ is an explanation for the slightly deeper trapping potential $\Phi\approx -0.70$ in figure \ref{fig:VISIT} e) when compared to figure \ref{fig:VISIT} b), despite the downramp which leads to a further decreased wave number-drive bunch length product away from its optimal value of $k_\mathrm{p} \sigma_z \approx \sqrt{2}$. This advantageous effect is confirmed by separate simulation runs where the drive beam energy is set to a very high value to preclude plasma lensing and to increase the betatron length. In this case $\tilde{Q}$ simply decreases from $\tilde{Q}(n_\mathrm{i})\approx 1.84$ before the ramp to $\tilde{Q}(n_\mathrm{f})\approx 1.50$ after the ramp, and both the worse $k_\mathrm{p} \sigma_z$ and the decreased $\tilde{Q}$ lead to reduced $\Phi \approx -0.59$ and reduced electric longitudinal field  $E_z \approx -7\ \mathrm{GV/m}$, compared to  $\Phi \approx -0.70$ and $E_z \approx -12.2\ \mathrm{GV/m}$ in case of the plasma lensing and scalloping, as shown in snapshots \ref{fig:VISIT} e) and f), respectively. 

This analysis demonstrates vividly the effect of stalling the plasma wave phase velocity $v_\mathrm{ph}$ and the possibility of inducing trapping even in unoptimized underdense photocathode cases. At the same time, in the blowout cavity behind the driver bunch no LIT electrons are injected by downramp injection, and  shallow trapping potentials suppress LIT electron trapping outside of the density downramp region.
Both effects combined allow for a clean underdense photocathode process and ultrahigh witness beam quality, unspoiled by dark current.

\begin{figure}[htbp]
 \hspace*{-0.015\textwidth}
\includegraphics[width=0.5\textwidth]{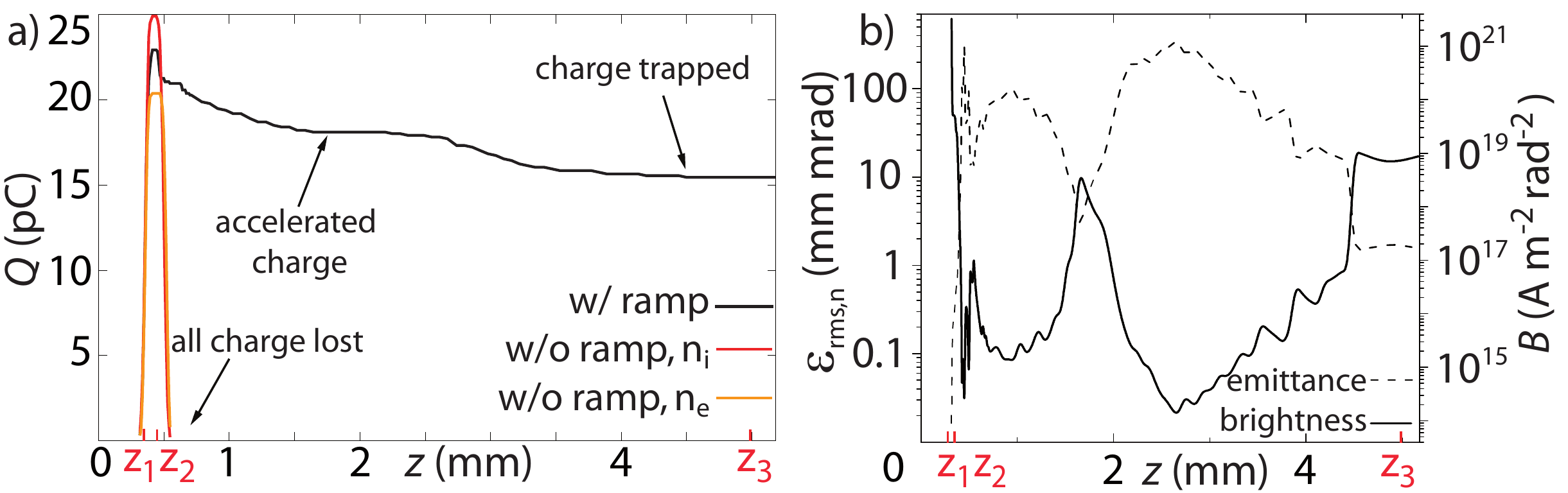}
\caption{Released He electron evolution. a) charge evolution for the DTH case (black) and cases without downramp (red and orange), in which case there is no trapping, and b) normalized emittance $\epsilon_\mathrm{n}$ and brightness $B$ of the DTH-generated witness bunch.}
\label{fig:downramp_no_downramp_emittance_brightness}
\end{figure}
Figure \ref{fig:downramp_no_downramp_emittance_brightness} a) confirms  that no trapping occurs when no downramp is present, based on simulation runs at high ($n_\mathrm{i}=1.5\times 10^{17}\ \mathrm{cm}^{-3}$, red plot) and low ($n_\mathrm{f}=0.6\times 10^{17}\ \mathrm{cm}^{-3}$, orange plot) densities, corresponding to the densities before and after the ramp. A total He  electron charge of $N_\mathrm{b,tot}e \approx 20-25\ \mathrm{pC}$ is generated, which is in very good agreement with a numero-analytical estimate of the charge yield based on Gaussian laser beam propagation and tunnel ionization. However, none of these electrons are trapped, but quickly slip out of the blowout and the co-propagating simulation box and are lost. In the DTH case (black plot), in contrast, 
most of the released charge is sufficiently  accelerated due to the downramp, and while even in this case substantial charge is subsequently lost, eventually $Q \approx 15.5\ \mathrm{pC}$ are fully trapped. 
Figure \ref{fig:downramp_no_downramp_emittance_brightness} b) shows the evolution of the witness bunch normalized emittance $\epsilon_\mathrm{n}$ (left $y$-axis, dashed line) and the corresponding brightness $B$ (right  $y$-axis, solid line). All electrons are counted, including those that are ultimately not trapped, which, in addition to mixing processes \cite{YunfengPhysRevSTABshort,PhysRevLett.112.035003Xu2014}, is responsible for the characteristic shape of the plots. The relatively sudden decrease of brightness and increase of emittance, respectively, at $z \approx 4.5\ \mathrm{mm}$ result from  the last untrapped electrons leaving the simulation box as the simulation progresses. The final values are  
$\epsilon_{\mathrm{n,rms}}\approx 2\times10^{-8}\ \mathrm{m\ rad}$ and $B \approx 9\times10^{18}\ \mathrm{A}\mathrm{m}^{-2}\mathrm{rad}^{-2}$, which is comparable to  
results from straightforward underdense photocathode simulations \cite{YunfengPhysRevSTABshort}. The final  witness bunch length is somewhat increased compared to normal Trojan Horse because of  plasma blowout lengthening, in this case to $\sigma_z \approx 9\ \mathrm{\mu m}$, leading to a peak witness current of $I_\mathrm{w} \approx 1\ \mathrm{kA}$. 
This bunch stretching in DTH and the effective suppression of dark current are advantages for FEL applications in the context of the radiation slippage, cooperation length and slice energy spread. 

In summary, the analysis of the DTH scheme shows that the implementation of experimentally viable downramps, such as  used in LWFA experiments, can dramatically increase the number of accelerator systems that qualify for  implementing  underdense photocathodes, for example aiming at driver-witness type emittance/brightness/luminosity transformer systems. Here, a complementary approach when compared to increasing the driver bunch current is used, namely shaping the plasma properties to stall the phase velocity to facilitate trapping. It is shown that trapping is possible already with driver currents of $I_\mathrm{d} \approx 3\ \mathrm{kA}$, thus reducing the otherwise required peak driver current by a factor of more than 2. This explicitly also makes this scheme accessible to double-hybrid schemes where the electron drive beam is generated via LWFA, as proposed for extending the energy gain in \cite{HiddingPRL2010PhysRevLett.104.195002short}. Recent experiments indicate that LWFA accelerated electron bunches are able to drive a PWFA blowout \cite{2014arXivMassonLabordeLWFAtoPWFAshort}. Such all-optical systems  offer advantages of compactness, and an inherent optimal synchronization between the underdense photocathode laser and the electron bunch driver, and further for seeding in the context of FEL, for other light sources and  for pump-probe type investigations in general. 
\begin{acknowledgments}
This work was supported by DFG, STFC 4070022104, DOE DE-SC0009533, DE-FG02-07ER46272, DE-FG03-92ER40693, by ONR N00014-06-1-0925 and Helmholtz VH-VI-503. We
acknowledge the assistance of the VSim development team. This research used computational resources of the
National Energy Research Scientific Computing Center, which is supported by DOE DE-AC02-05CH11231, and of
JUROPA, and of HLRN. D.A.J. acknowledges support of the UK EPSRC (EP/J018171/1) and the EC’s 7th
Framework Programme (LASERLAB-EUROPE no. 284464, EUCARD-2 project no. 312453) and the Extreme Light Infrastructure (ELI).
\end{acknowledgments}
\bibliography{Downramp_Assisted_Underdense_Photocathode}

\begin{thebibliography}{39}%
\makeatletter
\providecommand \@ifxundefined [1]{%
 \@ifx{#1\undefined}
}%
\providecommand \@ifnum [1]{%
 \ifnum #1\expandafter \@firstoftwo
 \else \expandafter \@secondoftwo
 \fi
}%
\providecommand \@ifx [1]{%
 \ifx #1\expandafter \@firstoftwo
 \else \expandafter \@secondoftwo
 \fi
}%
\providecommand \natexlab [1]{#1}%
\providecommand \enquote  [1]{``#1''}%
\providecommand \bibnamefont  [1]{#1}%
\providecommand \bibfnamefont [1]{#1}%
\providecommand \citenamefont [1]{#1}%
\providecommand \href@noop [0]{\@secondoftwo}%
\providecommand \href [0]{\begingroup \@sanitize@url \@href}%
\providecommand \@href[1]{\@@startlink{#1}\@@href}%
\providecommand \@@href[1]{\endgroup#1\@@endlink}%
\providecommand \@sanitize@url [0]{\catcode `\\12\catcode `\$12\catcode
  `\&12\catcode `\#12\catcode `\^12\catcode `\_12\catcode `\%12\relax}%
\providecommand \@@startlink[1]{}%
\providecommand \@@endlink[0]{}%
\providecommand \url  [0]{\begingroup\@sanitize@url \@url }%
\providecommand \@url [1]{\endgroup\@href {#1}{\urlprefix }}%
\providecommand \urlprefix  [0]{URL }%
\providecommand \Eprint [0]{\href }%
\providecommand \doibase [0]{http://dx.doi.org/}%
\providecommand \selectlanguage [0]{\@gobble}%
\providecommand \bibinfo  [0]{\@secondoftwo}%
\providecommand \bibfield  [0]{\@secondoftwo}%
\providecommand \translation [1]{[#1]}%
\providecommand \BibitemOpen [0]{}%
\providecommand \bibitemStop [0]{}%
\providecommand \bibitemNoStop [0]{.\EOS\space}%
\providecommand \EOS [0]{\spacefactor3000\relax}%
\providecommand \BibitemShut  [1]{\csname bibitem#1\endcsname}%
\let\auto@bib@innerbib\@empty
\bibitem [{\citenamefont {Akhiezer}\ and\ \citenamefont
  {Fainberg}(2008)}]{AkhiezerFainberg1949}%
  \BibitemOpen
  \bibfield  {author} {\bibinfo {author} {\bibfnamefont {A.}~\bibnamefont
  {Akhiezer}}\ and\ \bibinfo {author} {\bibfnamefont {Y.~B.}\ \bibnamefont
  {Fainberg}},\ }\href@noop {} {\bibfield  {journal} {\bibinfo  {journal} {Ukr.
  J. Phys.}\ }\textbf {\bibinfo {volume} {53 Special Issue}},\ \bibinfo {pages}
  {87} (\bibinfo {year} {2008})}\BibitemShut {NoStop}%
\bibitem [{\citenamefont {Fainberg}(1956)}]{Fainberg1956}%
  \BibitemOpen
  \bibfield  {author} {\bibinfo {author} {\bibfnamefont {Y.~B.}\ \bibnamefont
  {Fainberg}}\ }(\bibinfo  {publisher} {AIP},\ \bibinfo {address} {Geneva},\
  \bibinfo {year} {1956})\ p.~\bibinfo {pages} {84}\BibitemShut {NoStop}%
\bibitem [{\citenamefont {Veksler}(1956)}]{Veksler1956}%
  \BibitemOpen
  \bibfield  {author} {\bibinfo {author} {\bibfnamefont {V.}~\bibnamefont
  {Veksler}}\ }(\bibinfo  {publisher} {AIP},\ \bibinfo {address} {Geneva},\
  \bibinfo {year} {1956})\ p.~\bibinfo {pages} {80}\BibitemShut {NoStop}%
\bibitem [{\citenamefont {Budker}(1956)}]{Budker1956}%
  \BibitemOpen
  \bibfield  {author} {\bibinfo {author} {\bibfnamefont {G.}~\bibnamefont
  {Budker}}\ }(\bibinfo  {publisher} {AIP},\ \bibinfo {address} {Geneva},\
  \bibinfo {year} {1956})\ p.~\bibinfo {pages} {68}\BibitemShut {NoStop}%
\bibitem [{\citenamefont {Dawson}(1959)}]{Dawson1959PhysRev.113.383}%
  \BibitemOpen
  \bibfield  {author} {\bibinfo {author} {\bibfnamefont {J.~M.}\ \bibnamefont
  {Dawson}},\ }\href {\doibase 10.1103/PhysRev.113.383} {\bibfield  {journal}
  {\bibinfo  {journal} {Phys. Rev.}\ }\textbf {\bibinfo {volume} {113}},\
  \bibinfo {pages} {383} (\bibinfo {year} {1959})}\BibitemShut {NoStop}%
\bibitem [{\citenamefont {Akhiezer}\ and\ \citenamefont
  {Polovin}(1956)}]{AkhiezerPolovinJETP1956}%
  \BibitemOpen
  \bibfield  {author} {\bibinfo {author} {\bibfnamefont {A.~I.}\ \bibnamefont
  {Akhiezer}}\ and\ \bibinfo {author} {\bibfnamefont {R.~V.}\ \bibnamefont
  {Polovin}},\ }\href@noop {} {\bibfield  {journal} {\bibinfo  {journal}
  {JETP}\ }\textbf {\bibinfo {volume} {3}},\ \bibinfo {pages} {696} (\bibinfo
  {year} {1956})}\BibitemShut {NoStop}%
\bibitem [{\citenamefont {Esarey}\ \emph {et~al.}(1996)\citenamefont {Esarey},
  \citenamefont {Sprangle}, \citenamefont {Krall},\ and\ \citenamefont
  {Ting}}]{Esarey1996}%
  \BibitemOpen
  \bibfield  {author} {\bibinfo {author} {\bibfnamefont {E.}~\bibnamefont
  {Esarey}}, \bibinfo {author} {\bibfnamefont {P.}~\bibnamefont {Sprangle}},
  \bibinfo {author} {\bibfnamefont {J.}~\bibnamefont {Krall}}, \ and\ \bibinfo
  {author} {\bibfnamefont {A.}~\bibnamefont {Ting}},\ }\href@noop {} {\bibfield
   {journal} {\bibinfo  {journal} {IEEE Transactions on Plasma Science}\
  }\textbf {\bibinfo {volume} {24}},\ \bibinfo {pages} {252} (\bibinfo {year}
  {1996})}\BibitemShut {NoStop}%
\bibitem [{\citenamefont {Blumenfeld}\ \emph {et~al.}(2007)\citenamefont
  {Blumenfeld} \emph {et~al.}}]{Blumenfeld2007}%
  \BibitemOpen
  \bibfield  {author} {\bibinfo {author} {\bibfnamefont {I.}~\bibnamefont
  {Blumenfeld}} \emph {et~al.},\ }\href@noop {} {\bibfield  {journal} {\bibinfo
   {journal} {Nature}\ }\textbf {\bibinfo {volume} {445}},\ \bibinfo {pages}
  {741} (\bibinfo {year} {2007})}\BibitemShut {NoStop}%
\bibitem [{\citenamefont {Litos}\ \emph {et~al.}(2014)\citenamefont {Litos}
  \emph {et~al.}}]{Litos2014Natureshort}%
  \BibitemOpen
  \bibfield  {author} {\bibinfo {author} {\bibfnamefont {M.}~\bibnamefont
  {Litos}} \emph {et~al.},\ }\href@noop {} {\bibfield  {journal} {\bibinfo
  {journal} {Nature}\ }\textbf {\bibinfo {volume} {515}},\ \bibinfo {pages}
  {92} (\bibinfo {year} {2014})}\BibitemShut {NoStop}%
\bibitem [{\citenamefont {Hidding}\ \emph {et~al.}(2011)\citenamefont {Hidding}
  \emph {et~al.}}]{hiddingpatent2011short}%
  \BibitemOpen
  \bibfield  {author} {\bibinfo {author} {\bibfnamefont {B.}~\bibnamefont
  {Hidding}} \emph {et~al.},\ }\href@noop {} {\enquote {\bibinfo {title}
  {Method for generating electron beams in a hybrid plasma accelerator},}\ }
  (\bibinfo {year} {2011}),\ \bibinfo {note} {german Patent DE 10 2011 104
  858.1, US/PCT patent Ser. No. PCT/US12/043002}\BibitemShut {NoStop}%
\bibitem [{\citenamefont {Hidding}\ \emph
  {et~al.}(2012{\natexlab{a}})\citenamefont {Hidding} \emph
  {et~al.}}]{Hidding2012}%
  \BibitemOpen
  \bibfield  {author} {\bibinfo {author} {\bibfnamefont {B.}~\bibnamefont
  {Hidding}} \emph {et~al.},\ }\href {\doibase 10.1103/PhysRevLett.108.035001}
  {\bibfield  {journal} {\bibinfo  {journal} {Phys. Rev. Lett.}\ }\textbf
  {\bibinfo {volume} {108}},\ \bibinfo {pages} {035001} (\bibinfo {year}
  {2012}{\natexlab{a}})}\BibitemShut {NoStop}%
\bibitem [{\citenamefont {Hidding}\ \emph
  {et~al.}(2012{\natexlab{b}})\citenamefont {Hidding} \emph
  {et~al.}}]{hidding570beyondshort}%
  \BibitemOpen
  \bibfield  {author} {\bibinfo {author} {\bibfnamefont {B.}~\bibnamefont
  {Hidding}} \emph {et~al.},\ }\href {\doibase 10.1063/1.4773760} {\bibfield
  {journal} {\bibinfo  {journal} {AIP Conference Proceedings}\ }\textbf
  {\bibinfo {volume} {1507}},\ \bibinfo {pages} {570} (\bibinfo {year}
  {2012}{\natexlab{b}})}\BibitemShut {NoStop}%
\bibitem [{\citenamefont {Xi}\ \emph {et~al.}(2013)\citenamefont {Xi} \emph
  {et~al.}}]{YunfengPhysRevSTABshort}%
  \BibitemOpen
  \bibfield  {author} {\bibinfo {author} {\bibfnamefont {Y.}~\bibnamefont {Xi}}
  \emph {et~al.},\ }\href {\doibase 10.1103/PhysRevSTAB.16.031303} {\bibfield
  {journal} {\bibinfo  {journal} {Phys. Rev. ST Accel. Beams}\ }\textbf
  {\bibinfo {volume} {16}},\ \bibinfo {pages} {031303} (\bibinfo {year}
  {2013})}\BibitemShut {NoStop}%
\bibitem [{\citenamefont {Li}\ \emph {et~al.}(2013)\citenamefont {Li} \emph
  {et~al.}}]{LiPRL2013short}%
  \BibitemOpen
  \bibfield  {author} {\bibinfo {author} {\bibfnamefont {F.}~\bibnamefont {Li}}
  \emph {et~al.},\ }\href {\doibase 10.1103/PhysRevLett.111.015003} {\bibfield
  {journal} {\bibinfo  {journal} {Phys. Rev. Lett.}\ }\textbf {\bibinfo
  {volume} {111}},\ \bibinfo {pages} {015003} (\bibinfo {year}
  {2013})}\BibitemShut {NoStop}%
\bibitem [{\citenamefont {Xu}\ \emph {et~al.}(2014)\citenamefont {Xu} \emph
  {et~al.}}]{PhysRevLett.112.035003Xu2014}%
  \BibitemOpen
  \bibfield  {author} {\bibinfo {author} {\bibfnamefont {X.~L.}\ \bibnamefont
  {Xu}} \emph {et~al.},\ }\href {\doibase 10.1103/PhysRevLett.112.035003}
  {\bibfield  {journal} {\bibinfo  {journal} {Phys. Rev. Lett.}\ }\textbf
  {\bibinfo {volume} {112}},\ \bibinfo {pages} {035003} (\bibinfo {year}
  {2014})}\BibitemShut {NoStop}%
\bibitem [{\citenamefont {Chen}\ \emph {et~al.}(1985)\citenamefont {Chen} \emph
  {et~al.}}]{Chen1985PhysRevLettshort}%
  \BibitemOpen
  \bibfield  {author} {\bibinfo {author} {\bibfnamefont {P.}~\bibnamefont
  {Chen}} \emph {et~al.},\ }\href {\doibase 10.1103/PhysRevLett.54.693}
  {\bibfield  {journal} {\bibinfo  {journal} {Phys. Rev. Lett.}\ }\textbf
  {\bibinfo {volume} {54}},\ \bibinfo {pages} {693} (\bibinfo {year}
  {1985})}\BibitemShut {NoStop}%
\bibitem [{\citenamefont {Rosenzweig}\ \emph {et~al.}(1988)\citenamefont
  {Rosenzweig} \emph {et~al.}}]{Rosenzweig1988PRLshort}%
  \BibitemOpen
  \bibfield  {author} {\bibinfo {author} {\bibfnamefont {J.~B.}\ \bibnamefont
  {Rosenzweig}} \emph {et~al.},\ }\href {\doibase 10.1103/PhysRevLett.61.98}
  {\bibfield  {journal} {\bibinfo  {journal} {Phys. Rev. Lett.}\ }\textbf
  {\bibinfo {volume} {61}},\ \bibinfo {pages} {98} (\bibinfo {year}
  {1988})}\BibitemShut {NoStop}%
\bibitem [{\citenamefont {Tajima}\ and\ \citenamefont
  {Dawson}(1979)}]{Tajima1979}%
  \BibitemOpen
  \bibfield  {author} {\bibinfo {author} {\bibfnamefont {T.}~\bibnamefont
  {Tajima}}\ and\ \bibinfo {author} {\bibfnamefont {J.~M.}\ \bibnamefont
  {Dawson}},\ }\href@noop {} {\bibfield  {journal} {\bibinfo  {journal}
  {Physical Review Letters}\ }\textbf {\bibinfo {volume} {43}},\ \bibinfo
  {pages} {267} (\bibinfo {year} {1979})}\BibitemShut {NoStop}%
\bibitem [{\citenamefont {Pukhov}\ and\ \citenamefont {Meyer-ter
  Vehn}(2002)}]{Pukhov2002}%
  \BibitemOpen
  \bibfield  {author} {\bibinfo {author} {\bibfnamefont {A.}~\bibnamefont
  {Pukhov}}\ and\ \bibinfo {author} {\bibfnamefont {J.}~\bibnamefont {Meyer-ter
  Vehn}},\ }\href@noop {} {\bibfield  {journal} {\bibinfo  {journal} {Applied
  Physics B-Lasers and Optics}\ }\textbf {\bibinfo {volume} {74}},\ \bibinfo
  {pages} {355} (\bibinfo {year} {2002})}\BibitemShut {NoStop}%
\bibitem [{\citenamefont {Kostyukov}\ \emph {et~al.}(2009)\citenamefont
  {Kostyukov} \emph {et~al.}}]{KostyukovTrappingPRL2009short}%
  \BibitemOpen
  \bibfield  {author} {\bibinfo {author} {\bibfnamefont {I.}~\bibnamefont
  {Kostyukov}} \emph {et~al.},\ }\href {\doibase
  10.1103/PhysRevLett.103.175003} {\bibfield  {journal} {\bibinfo  {journal}
  {Phys. Rev. Lett.}\ }\textbf {\bibinfo {volume} {103}},\ \bibinfo {pages}
  {175003} (\bibinfo {year} {2009})}\BibitemShut {NoStop}%
\bibitem [{\citenamefont {Kirby}\ \emph {et~al.}(2009)\citenamefont {Kirby}
  \emph {et~al.}}]{Kirby2009PRSTshort}%
  \BibitemOpen
  \bibfield  {author} {\bibinfo {author} {\bibnamefont {Kirby}} \emph
  {et~al.},\ }\href {\doibase 10.1103/PhysRevSTAB.12.051302} {\bibfield
  {journal} {\bibinfo  {journal} {Phys. Rev. ST Accel. Beams}\ }\textbf
  {\bibinfo {volume} {12}},\ \bibinfo {pages} {051302} (\bibinfo {year}
  {2009})}\BibitemShut {NoStop}%
\bibitem [{\citenamefont {Bulanov}\ \emph {et~al.}(1998)\citenamefont {Bulanov}
  \emph {et~al.}}]{BulanovDensityTransition1998PhysRevEshort}%
  \BibitemOpen
  \bibfield  {author} {\bibinfo {author} {\bibfnamefont {S.}~\bibnamefont
  {Bulanov}} \emph {et~al.},\ }\href {\doibase 10.1103/PhysRevE.58.R5257}
  {\bibfield  {journal} {\bibinfo  {journal} {Phys. Rev. E}\ }\textbf {\bibinfo
  {volume} {58}},\ \bibinfo {pages} {R5257} (\bibinfo {year}
  {1998})}\BibitemShut {NoStop}%
\bibitem [{\citenamefont {Geddes}\ \emph {et~al.}(2008)\citenamefont {Geddes}
  \emph {et~al.}}]{GeddesDensityDownrampPRL2008short}%
  \BibitemOpen
  \bibfield  {author} {\bibinfo {author} {\bibfnamefont {C.~G.~R.}\
  \bibnamefont {Geddes}} \emph {et~al.},\ }\href {\doibase
  10.1103/PhysRevLett.100.215004} {\bibfield  {journal} {\bibinfo  {journal}
  {Phys. Rev. Lett.}\ }\textbf {\bibinfo {volume} {100}},\ \bibinfo {pages}
  {215004} (\bibinfo {year} {2008})}\BibitemShut {NoStop}%
\bibitem [{\citenamefont {Schmid}\ \emph {et~al.}(2010)\citenamefont {Schmid}
  \emph {et~al.}}]{SchmidDensityTransPhysRevSTABshort}%
  \BibitemOpen
  \bibfield  {author} {\bibinfo {author} {\bibfnamefont {K.}~\bibnamefont
  {Schmid}} \emph {et~al.},\ }\href {\doibase 10.1103/PhysRevSTAB.13.091301}
  {\bibfield  {journal} {\bibinfo  {journal} {Phys. Rev. ST Accel. Beams}\
  }\textbf {\bibinfo {volume} {13}},\ \bibinfo {pages} {091301} (\bibinfo
  {year} {2010})}\BibitemShut {NoStop}%
\bibitem [{\citenamefont {Gonsalves}\ \emph {et~al.}(2011)\citenamefont
  {Gonsalves} \emph {et~al.}}]{GonsalvesNatPhys2011short}%
  \BibitemOpen
  \bibfield  {author} {\bibinfo {author} {\bibnamefont {Gonsalves}} \emph
  {et~al.},\ }\href {http://dx.doi.org/10.1038/nphys2071} {\bibfield  {journal}
  {\bibinfo  {journal} {Nat Phys}\ }\textbf {\bibinfo {volume} {7}},\
  (\bibinfo {year} {2011})}\BibitemShut {NoStop}%
\bibitem [{\citenamefont {Suk}\ \emph {et~al.}(2001)\citenamefont {Suk} \emph
  {et~al.}}]{SukRosenzweigDensityTransition2001PhysRevLettshort}%
  \BibitemOpen
  \bibfield  {author} {\bibinfo {author} {\bibfnamefont {H.}~\bibnamefont
  {Suk}} \emph {et~al.},\ }\href {\doibase 10.1103/PhysRevLett.86.1011}
  {\bibfield  {journal} {\bibinfo  {journal} {Phys. Rev. Lett.}\ }\textbf
  {\bibinfo {volume} {86}},\ \bibinfo {pages} {1011} (\bibinfo {year}
  {2001})}\BibitemShut {NoStop}%
\bibitem [{\citenamefont {Wiggins}\ \emph {et~al.}(2010)\citenamefont {Wiggins}
  \emph {et~al.}}]{Wiggins1percent2010PPCFshort}%
  \BibitemOpen
  \bibfield  {author} {\bibinfo {author} {\bibfnamefont {S.~M.}\ \bibnamefont
  {Wiggins}} \emph {et~al.},\ }\href
  {http://stacks.iop.org/0741-3335/52/i=12/a=124032} {\bibfield  {journal}
  {\bibinfo  {journal} {Plasma Physics and Controlled Fusion}\ }\textbf
  {\bibinfo {volume} {52}},\ \bibinfo {pages} {124032} (\bibinfo {year}
  {2010})}\BibitemShut {NoStop}%
\bibitem [{\citenamefont {Lundh}\ \emph {et~al.}(2011)\citenamefont {Lundh}
  \emph {et~al.}}]{LundhKiloampsNatPhys2011short}%
  \BibitemOpen
  \bibfield  {author} {\bibinfo {author} {\bibfnamefont {O.}~\bibnamefont
  {Lundh}} \emph {et~al.},\ }\href {http://dx.doi.org/10.1038/nphys1872}
  {\bibfield  {journal} {\bibinfo  {journal} {Nat Phys}\ }\textbf {\bibinfo
  {volume} {7}},\ \bibinfo {pages} {219} (\bibinfo {year} {2011})}\BibitemShut
  {NoStop}%
\bibitem [{\citenamefont {Nieter}\ and\ \citenamefont
  {Cary}(2004)}]{Nieter2004448}%
  \BibitemOpen
  \bibfield  {author} {\bibinfo {author} {\bibfnamefont {C.}~\bibnamefont
  {Nieter}}\ and\ \bibinfo {author} {\bibfnamefont {J.~R.}\ \bibnamefont
  {Cary}},\ }\href {\doibase DOI: 10.1016/j.jcp.2003.11.004} {\bibfield
  {journal} {\bibinfo  {journal} {Journal of Computational Physics}\ }\textbf
  {\bibinfo {volume} {196}},\ \bibinfo {pages} {448 } (\bibinfo {year}
  {2004})}\BibitemShut {NoStop}%
\bibitem [{\citenamefont {Umstadter}\ \emph {et~al.}(1995)\citenamefont
  {Umstadter}, \citenamefont {Kim},\ and\ \citenamefont
  {Dodd}}]{Umstadterpatent1995}%
  \BibitemOpen
  \bibfield  {author} {\bibinfo {author} {\bibfnamefont {D.}~\bibnamefont
  {Umstadter}}, \bibinfo {author} {\bibfnamefont {J.-K.}\ \bibnamefont {Kim}},
  \ and\ \bibinfo {author} {\bibfnamefont {E.}~\bibnamefont {Dodd}},\
  }\href@noop {} {\enquote {\bibinfo {title} {Method and apparatus for
  generating and accelerating ultrashort electron pulses},}\ } (\bibinfo {year}
  {1995}),\ \bibinfo {note} {{US} patent Ser. No. 5,789,876}\BibitemShut
  {NoStop}%
\bibitem [{\citenamefont {Umstadter}\ \emph {et~al.}(1996)\citenamefont
  {Umstadter}, \citenamefont {Kim},\ and\ \citenamefont
  {Dodd}}]{Umstadter1996CollidingPulsesPhysRevLett.76.2073}%
  \BibitemOpen
  \bibfield  {author} {\bibinfo {author} {\bibfnamefont {D.}~\bibnamefont
  {Umstadter}}, \bibinfo {author} {\bibfnamefont {J.~K.}\ \bibnamefont {Kim}},
  \ and\ \bibinfo {author} {\bibfnamefont {E.}~\bibnamefont {Dodd}},\ }\href
  {\doibase 10.1103/PhysRevLett.76.2073} {\bibfield  {journal} {\bibinfo
  {journal} {Phys. Rev. Lett.}\ }\textbf {\bibinfo {volume} {76}},\ \bibinfo
  {pages} {2073} (\bibinfo {year} {1996})}\BibitemShut {NoStop}%
\bibitem [{\citenamefont {Chen}\ \emph {et~al.}(2006)\citenamefont {Chen} \emph
  {et~al.}}]{ChenJAP200610short}%
  \BibitemOpen
  \bibfield  {author} {\bibinfo {author} {\bibfnamefont {M.}~\bibnamefont
  {Chen}} \emph {et~al.},\ }\href {\doibase DOI:10.1063/1.2179194} {\bibfield
  {journal} {\bibinfo  {journal} {Journal of Applied Physics}\ }\textbf
  {\bibinfo {volume} {99}},\ \bibinfo {pages} {056109} (\bibinfo {year}
  {2006})}\BibitemShut {NoStop}%
\bibitem [{\citenamefont {Bruhwiler}\ \emph {et~al.}(2003)\citenamefont
  {Bruhwiler} \emph {et~al.}}]{Bruhwiler2003short}%
  \BibitemOpen
  \bibfield  {author} {\bibinfo {author} {\bibfnamefont {D.~L.}\ \bibnamefont
  {Bruhwiler}} \emph {et~al.},\ }\href {\doibase 10.1063/1.1566027} {\bibfield
  {journal} {\bibinfo  {journal} {Physics of Plasmas}\ }\textbf {\bibinfo
  {volume} {10}},\ \bibinfo {pages} {2022} (\bibinfo {year}
  {2003})}\BibitemShut {NoStop}%
\bibitem [{\citenamefont {Barov}\ \emph {et~al.}(2004)\citenamefont {Barov}
  \emph {et~al.}}]{BarovEnergyLossQtildePRST2004short}%
  \BibitemOpen
  \bibfield  {author} {\bibinfo {author} {\bibfnamefont {N.}~\bibnamefont
  {Barov}} \emph {et~al.},\ }\href {\doibase 10.1103/PhysRevSTAB.7.061301}
  {\bibfield  {journal} {\bibinfo  {journal} {Phys. Rev. ST Accel. Beams}\
  }\textbf {\bibinfo {volume} {7}},\ \bibinfo {pages} {061301} (\bibinfo {year}
  {2004})}\BibitemShut {NoStop}%
\bibitem [{\citenamefont {Fubiani}\ \emph {et~al.}(2006)\citenamefont {Fubiani}
  \emph {et~al.}}]{PhysRevEFubianiDensityGradientsshort}%
  \BibitemOpen
  \bibfield  {author} {\bibinfo {author} {\bibfnamefont {G.}~\bibnamefont
  {Fubiani}} \emph {et~al.},\ }\href {\doibase 10.1103/PhysRevE.73.026402}
  {\bibfield  {journal} {\bibinfo  {journal} {Phys. Rev. E}\ }\textbf {\bibinfo
  {volume} {73}},\ \bibinfo {pages} {026402} (\bibinfo {year}
  {2006})}\BibitemShut {NoStop}%
\bibitem [{\citenamefont {Buck}\ \emph {et~al.}(2013)\citenamefont {Buck} \emph
  {et~al.}}]{BuckShockFrontInjector2013short}%
  \BibitemOpen
  \bibfield  {author} {\bibinfo {author} {\bibfnamefont {A.}~\bibnamefont
  {Buck}} \emph {et~al.},\ }\href {\doibase 10.1103/PhysRevLett.110.185006}
  {\bibfield  {journal} {\bibinfo  {journal} {Phys. Rev. Lett.}\ }\textbf
  {\bibinfo {volume} {110}},\ \bibinfo {pages} {185006} (\bibinfo {year}
  {2013})}\BibitemShut {NoStop}%
\bibitem [{\citenamefont {Pak}\ \emph {et~al.}(2010)\citenamefont {Pak} \emph
  {et~al.}}]{PakPRL2010PhysRevLettshort}%
  \BibitemOpen
  \bibfield  {author} {\bibinfo {author} {\bibfnamefont {A.}~\bibnamefont
  {Pak}} \emph {et~al.},\ }\href {\doibase 10.1103/PhysRevLett.104.025003}
  {\bibfield  {journal} {\bibinfo  {journal} {Phys. Rev. Lett.}\ }\textbf
  {\bibinfo {volume} {104}},\ \bibinfo {pages} {025003} (\bibinfo {year}
  {2010})}\BibitemShut {NoStop}%
\bibitem [{\citenamefont {Hidding}\ \emph {et~al.}(2010)\citenamefont {Hidding}
  \emph {et~al.}}]{HiddingPRL2010PhysRevLett.104.195002short}%
  \BibitemOpen
  \bibfield  {author} {\bibinfo {author} {\bibfnamefont {B.}~\bibnamefont
  {Hidding}} \emph {et~al.},\ }\href {\doibase 10.1103/PhysRevLett.104.195002}
  {\bibfield  {journal} {\bibinfo  {journal} {Phys. Rev. Lett.}\ }\textbf
  {\bibinfo {volume} {104}},\ \bibinfo {pages} {195002} (\bibinfo {year}
  {2010})}\BibitemShut {NoStop}%
\bibitem [{\citenamefont {{Masson-Laborde}}\ \emph {et~al.}(2014)\citenamefont
  {{Masson-Laborde}} \emph {et~al.}}]{2014arXivMassonLabordeLWFAtoPWFAshort}%
  \BibitemOpen
  \bibfield  {author} {\bibinfo {author} {\bibfnamefont {P.~E.}\ \bibnamefont
  {{Masson-Laborde}}} \emph {et~al.},\ }\href@noop {} {\bibfield  {journal}
  {\bibinfo  {journal} {ArXiv e-prints}\ } (\bibinfo {year} {2014})},\ \Eprint
  {http://arxiv.org/abs/1408.2494} {arXiv:1408.2494 [physics.plasm-ph]}
  \BibitemShut {NoStop}%
\end{thebibliography}%


%
\end{document}